\documentclass[secnumarabic, graphics,floatfix, nofootinbib,tightenlines,nobibnotes, aps, prb, twocolumn]{revtex4-1}
\usepackage{amsmath,amssymb}
\usepackage{graphicx}
\usepackage{dcolumn}
\usepackage{bm}
\usepackage{braket}
\usepackage{verbatim}
\usepackage{float}
\usepackage{bbold}
\usepackage{color}
\usepackage{xcolor}
\usepackage{relsize}

\usepackage{slashed}
\usepackage{amsthm}
\usepackage[colorlinks=true ,urlcolor=blue,urlbordercolor={0 1 1}]{hyperref}

\begin{document}

\title{Composite Fermion Insulator in Opposite-Fields Quantum Hall Bilayers}
\author{%
Ya-Hui Zhang
}
\email{yahuizh@mit.edu}
\affiliation{
Department of Physics, Massachusetts Institute of Technology, Cambridge, MA, USA
}

\date{\today}

\begin{abstract}
Recently several moir\'e super-lattice systems are proposed to host nearly flat $\pm$ Chern bands: the bands of the two valleys have opposite Chern numbers. In these systems the charge of each valley is separately conserved. For the $C=\pm 1$ case, in the perfect flat band limit, the system can be mapped to two Landau levels from opposite magnetic fields. Motivated by these promising experimental realizations, we consider a quantum Hall bilayers from opposite magnetic fields close to the filling $\nu_T=\frac{1}{2}+\frac{1}{2}$. We add inter-layer repulsive interaction starting from two decoupled Composite Fermion Liquids (CFL) with opposite chiralities. In this case physical exciton is frustrated from condensation, unlike the conventional quantum Hall bilayers. We argue that more natural phases are the exciton condensates between composite fermions or between slave bosons. The resulting states are insulators with neutral Fermi surfaces coupled to an emergent $U(1)$ gauge field without Chern-Simons term. This insulating state is a generalization of the well-known CFL state  and  may  potentially emerge in the moir\'e systems with $C=\pm 1$ narrow bands.  Finally we also comment on the possibility of  a topological superconductor from charge frustration in opposite-fields quantum Hall bilayers or in $C=\pm 1$ flat bands. 
\end{abstract}

\pacs{Valid PACS appear here}
\maketitle

\section{Introduction}
Recently several interesting strongly correlated states have been experimentally realized in moir\'e superlattices in twisted Van der Waals heterostructures. \cite{spanton2018observation,cao2018correlated,cao2018unconventional,chen2018gate,yankowitz2018tuning}. 
The potential of the moir\'e systems is apparently not limited to these existing examples. There are also proposals for nearly flat $\pm$ Chern bands in moir\'e super-lattices from twisting graphene layers or transition metal dichalcogenide (TMD) layers\cite{zhang2018moir,chittari2018gate,wu2018layer,zhang2018bridging}. In these systems valley can be viewed as a pseudo-spin and the bands of the two valleys have opposite Chern numbers. Strongly correlations combined with the non-trivial topology can lead to interesting new phases. One natural possibility is quantum-Hall like states through the spontaneous polarization of the valley as suggested by Ref.~\onlinecite{zhang2018moir}.  In this paper we explore  possible time-reversal invariant phases in the flat band limit. For simplicity we consider the case with Chern numbers $C=\pm 1$ for the two valleys and ignore the physical spin\footnote{Experimentally this corresponds to the case with large in-plane magnetic field.}. In this case the system can be mapped to two Landau levels from opposite magnetic fields.  This quantum Hall bilayers from opposite magnetic fields may host new phases beyond the well-studied conventional quantum Hall bilayers from the same magnetic field. We are going to explore a small subset  of the unexplored physics in this new set-ups and leave detailed studies of the whole phase diagram to the future. Landau level can be viewed as an ideal limit for a $C=1$ flat band. Therefore in most of the time we focus on the opposite-fields quantum Hall bilayers for simplicity. We will comment on the implications of our results for the experimentally more relevant $C=\pm 1$ flat bands systems.

We focus on the half filling of the opposite-fields quantum Hall bilayers: $\nu_T=\frac{1}{2}+\frac{1}{2}$. When the inter-layer distance $d$ is infinite, the ground state consists of two  decoupled  CFLs from the two layers. Then we add inter-layer repulsion by decreasing $d$. For the conventional quantum Hall bilayers, exciton condensation is  favored  because of inter-layer repulsion\cite{eisenstein2014exciton,moon1995spontaneous,yang1996spontaneous}.  Although exciton condensation between composite fermions is proposed theoretically at intermediate distance\cite{alicea2009interlayer,you2017interlayer}, there is no evidence for it in numerical simulations\cite{zhu2017numerical}. In contrast, in our new set-up, the physical exciton is frustrated from condensation. Therefore we can avoid the exciton condensation phase and explore more interesting possibilities. We argue that one natural possibility at intermediate distance is the exciton condensation between composite fermions or between slave-bosons, which results in an insulating state with neutral Fermi surfaces coupled to an internal $U(1)$ gauge field. We dub it as a composite fermion insulator (CFI) state. This CFI phase can be viewed as a generation of the familiar CFL phase.  The CFL state can be constructed in a parton construction of electron operator: $c=bf$. $f$ is a neutral fermion which forms a Fermi surface, while the charged boson $b$ is in the $\nu=\frac{1}{2}$ Laughlin state(or in a $U(1)_2$ topological ordered state). The proposed CFI state is simply making the boson $b$ part in a double semion state.  This CFI phase is different from the spinon Fermi surface state\cite{motrunich2005variational,lee2005u} in a spin model. The neutral fermion in the CFI phase does not need to carry any quantum number, unlike the spinon in the more familiar spin liquid phases. The neutral fermion here should be better viewed as a composite fermion which bind one electron and two quasi-holes with charge $-\frac{e}{2}$, similar to the composite fermion in the half-filled Landau level\cite{halperin1993theory}.  We describe the CFI phase starting from two CFL in both HLR and Dirac pictures. In the Dirac picture, the CFI phase has an explicit particle-hole symmetry, in addition to the time reversal symmetry.

In addition to the exotic CFI phase, we also suggest a quite novel mechanism for superconductivity in the opposite-fields quantum Hall bilayers. In this model inter-layer electron-electron pair does not fell any magnetic field and therefore can have a well-defined momentum. Though $s$ wave pairing is suppressed by the inter-layer repulsion, higher angular momentum cooper pair can be energetically favorable and can freely condense.

\section{CFI in the HLR Picture}
 At the filling $\nu_T=\frac{1}{2}+\frac{1}{2}$ of the quantum Hall bilayers with opposite magnetic fields, the decoupled limit have two CFLs with opposite chirality.  In this section we describe each CFL by the standard Halperin-Lee-Read (HLR) theory from flux attachment\cite{halperin1993theory}.  In this paper we use a modified version of the HLR theory for two reasons: (1) The coefficient of the Chern-Simons term should be an integer to be mathematically well-defined; (2) The composite fermions are neutral and should not couple to the physical gauge field $A$.  For each layer $I=1,2$, we do parton construction $c_I=b_I f_I$. Then we let the neutral fermion $f_I$ form a Fermi surface and let the charged boson $b_I$ go to the $\nu=\pm \frac{1}{2}$ Laughlin state. The low energy theory is:
\begin{align}
L=\sum_I L_{FS}[f_I,a_I]+L^b_{I}[b_I,A_I-a_I]
\label{eq:decouple_CFLs}
\end{align}
where $I=1,2$ is the layer index. $a_I$ is an $U(1)$ gauge field from the parton $c_I=b_I f_I$. $L_{FS}[f_I,a_I]$ describes a Fermi surface formed by the neutral composite fermion $f_I$ coupled to the $U(1)$ gauge field $a_I$. $L^b_{I}[b_I,A_I-a_I]$ is the theory for the $\nu=\pm \frac{1}{2}$ bosonic Laughlin state coupled to gauge field $A_I-a_I$:
\begin{equation}
  L^b_{I}[b_I,A_I-a_I]=\mp \frac{2}{4\pi}\alpha_I d \alpha_I +\frac{A_I-a_I}{2\pi} d \alpha_I
  \label{eq:double-semion}
\end{equation}
where $\mp$ corresponds to layer $1$ and $2$ separately. $\alpha_I$ is another gauge field to describe the $U(1)_2$ bosonic sector. The above action can be also understood as the double-semion state  $(2,-2,0)$ for the two component boson $(b_1,b_2)$. Integrating $\alpha_I$ recovers  the HLR theory for two decoupled CFLs. But to be more rigorous we will use the above theory as the starting point instead of the original version of the HLR theory.

The above action is invariant to the time reversal symmetry which interchanges the two layers:
\begin{equation}
  f_a(t,x)\rightarrow \tau^x_{ab}f_b(-t,x)
\end{equation}

\begin{align}
  &(a^1_0(t,x),\vec{a}^1(t,x))\rightarrow (a^2_0(-t,x),-\vec{a}^2(t,x))\notag\\
  &(a^2_0(t,x),\vec{a}^2(t,x))\rightarrow (a^1_0(-t,x),-\vec{a}^1(t,x))
\end{align}

\begin{align}
  &(\alpha^1_0(t,x),\vec{\alpha}^1(t,x))\rightarrow (-\alpha^2_0(-t,x),\vec{\alpha}^2(t,x))\notag\\
  &(\alpha^2_0(t,x),\vec{\alpha}^2(t,x))\rightarrow (-\alpha^1_0(-t,x),\vec{\alpha}^1(t,x))
\end{align}
and
\begin{align}
  &(A^1_0(t,x),\vec{A}^1(t,x))\rightarrow (A^2_0(-t,x),-\vec{A}^2(t,x))\notag\\
  &(A^2_0(t,x),\vec{A}^2(t,x))\rightarrow (A^1_0(-t,x),-\vec{A}^1(t,x))
\end{align}

Note that the time reversal transformation for the $\alpha_I$ gauge field looks like a $CT$ symmetry.

The system itself also has a particle hole (PH) symmetry which transforms electron of one layer to the hole of the same layer. However, in HLR picture it is hard to track the PH symmetry\footnote{It is still debated whether HLR theory is PH symmetric or not.}. We delay the discussion of PH symmetry to next section using Dirac picture for CFLs.

We can redefine $A_c=\frac{A_1+A_2}{2}$ and $A_s=\frac{A_1-A_2}{2}$. $A_c$ couples to the physical charge, while $A_s$ couples to the layer charge, which we view as $S_z$ of a pseudospin.  We also define $a_c=\frac{a_1+a_2}{2}$ and $a_s=\frac{a_1-a_2}{2}$. From the Laughlin argument\cite{laughlin1981quantized}, a $2\pi$ flux $d a_1$ carries charge $1/2$ of $b_1$, while $2\pi$ flux of $da_2$ carries $-1/2$ charge of $b_2$. Because of this non-trivial response of the bosonic double-semion sector, $d a_c$ carries a $S_z$ charge, while $d a_s$ carries a physical charge.

\subsection{Summary of several CFI Phases}

Starting from the decoupled two CFLs at the infinite distance separating the two layers ($d=\infty$), we can reduce the distance and discuss the possible phases.  Because of the non-trivial response of the bosonic state, magnetic flux $d a_c$ and $d a_s$ carry either $S_z$ or physical charge $Q$. In the tradtional quantum Hall bilayers, the adding intra-layer repulsion $V(\mathbf q)\rho_1(\mathbf q)\rho_2(-\mathbf q)$ is represented by $V(\mathbf q)(d a_c da_c -d a_s d a_s)$. Therefore $\mathbf{a_c}$ is suppressed while $\mathbf{a_s}$ is enhanced. $\mathbf{a_s}$ can mediates an inter-layer pairing through the ``Amperean Pairing'' mechanism, which then leads to the exciton condensation phase\cite{sodemann2017composite} at infinitesimal finite distance $d$.  In the quantum Hall bilayers with opposite magnetic field, the intra-layer repulsion $V(\mathbf q)\rho_1(\mathbf q)\rho_2(-\mathbf q)$ is represented by $V(\mathbf q)(-d a_c da_c +d a_s d a_s)$. In this case $\mathbf{a_s}$ is actually suppressed while $\mathbf{a_c}$ is enhanced. $\mathbf{a_c}$ generically suppresses the zero-momentum pairing\cite{metlitski2015cooper}. Therefore we expect that the state with two decoupled CFLs is stable to infinitesimal finite distance for this new set up.

In the intermediate distance it is natural to expect that the decoupled CFLs are unstable to another phase. Inter-layer repulsion naturally bound an electron in one layer with a hole in the other layer to a bosonic exciton. In the traditional case this bosonic exciton can then condense and the system is in a symmetry breaking state with an additional integer Hall conductivity. In our new model, the excitons still fell a doubly enlarged magnetic field and can not condense. Therefore we can avoid the exciton condensation phase in this new set up.

Instead of the physical exciton condensation, more natural possibilities are the exciton condensation between the partons $b_I$ or between $f_I$. There are three possibilities, which are summarized in Table.~\ref{table:CFI phases}.   All of these  phases are insulators which can show quantum oscillations under an external magnetic field $B_c=d A_c$. We dub these phases as composite fermion insulator (CFI). They are distinguished by their different responses to $A_s$. CFI3 is a superfluid under $A_s$. Both CFI1 and CFI2 are metals under $A_s$. However, the gapless charge of $A_s$ is carried by the internal flux $da$ in CFI1 and is carried by the composite fermions in CFI2. We are going to discuss these three CFI phases separately in the following sections.

\onecolumngrid

\begin{table}[h]
\centering
\begin{tabular}{c|c|c|c|c|c|c}

\hline
Phase&Mechanism& Low Energy Theory& Under $\mathbf{E_c}$ &Under $\mathbf{E_s}$&Under $B_c$&under $B_s$\\
\hline
CFI1& $\langle f_1^\dagger f_2 \rangle\neq 0$ & $L_{FS}[f_1,f_2,a]-\frac{1}{2\pi}A_s da +\frac{1}{4\pi}A_c d A_s+\frac{1}{4\pi} A_s d A_c$&Insulator&Metal&$da=\alpha B_c$&$\rho_f$ changes\\
\hline
CFI2& $\langle b_1^\dagger b_2 \rangle\neq 0$ & $L_{FS}[f_1,a+A_s]+L_{FS}[f_2,a-A_s]$&Insulator& Metal& $da=\alpha B_c$&Quantum Oscillations\\
\hline
CFI3& $\langle b_1^\dagger b_2 \rangle\neq 0; \langle f_1^\dagger f_2 \rangle \neq 0$ & $L_{FS}[f_1,f_2,a]+\rho_s A_s^2$&Insulator& Superfluid& $da=\alpha B_c$&Meissner Effect\\
\hline

\hline
\end{tabular}
\caption{A summary of three different CFI phases. The three classes are enriched by the $U(1)_s$ symmetry corresponding to $A_s$. }
\label{table:CFI phases}
\end{table}

\twocolumngrid

\subsection{CFI1: Exciton Condensation of  Fermions}

In the intermediate distance, composite fermions should also fell a residual short ranged inter-layer repulsion: $+f_1^\dagger f_1 f_2^\dagger f_2=-(f_1^\dagger f_2)(f_2^\dagger f_1)$. We can see that the exciton condensation between the composite fermions are favored by this inter-layer repulsion. The exciton condensation between composite fermions have been proposed in the traditional quantum Hall bilayers with the same magnetic field\cite{alicea2009interlayer,you2017interlayer}. However in the numerical simulation\cite{zhu2017numerical} exciton condensation is favored. The quantum Hall bilayers with opposite magnetic field is probably a better platform for composite fermion exciton condensation because the physical exciton condensation can be avoided. We propose the phase with $\langle f_1^\dagger f_2 \rangle \neq 0$ in the inter-mediate region and discuss its properties.  Such a bosonic exciton condensation apparently higgses the $a_s$ gauge field. We can then identify $a_1=a_2=a$. The final theory is:
\begin{align}
L_{CFI_1}=L_{FS}[f_1,f_2,a]+\sum_I L^b_{I}[b_I,A_I-a]
\end{align}
where
\begin{equation}
  L^b_I[b_I,A_I-a]=\mp \frac{2}{4\pi}\alpha_I d \alpha_I +\frac{A_I-a}{2\pi} d \alpha_I
\end{equation}

$L_{FS}[f_1,f_2,a]$ contain two Fermi surfaces coupled to the gauge field $a$. Because of the exciton condensation $\langle f_1^\dagger f_2 \rangle=\Delta \neq 0$, the areas of the two Fermi surfaces are not equal and depend on $\Delta$, though the sum is fixed.

We can also integrate $\alpha_I$ now.  Equivalently the phase can be described as:
\begin{align}
L_{CFI_1}=L_{FS}[f_1,f_2,a]-\frac{1}{2\pi}A_s da +\frac{1}{4\pi}A_c d A_s+\frac{1}{4\pi} A_s d A_c
\label{eq:cfi_low_energy}
\end{align}

Because of the $f_1^\dagger f_2$ term, $f_1$ and $f_2$ are mixed and we use $f$ to denote them. Both $f$ and monopole of $da$ are neutral. Therefore this phase is an insulator under $A_c$. The monopole of $2\pi$ flux for $\mathbf{a}$ carries  charge under $A_s$. Therefore this phase is metallic under $A_s$, similar to the behavior of the CFL under physical gauge field. The fermion $f$ can still be understood as a composite fermion. The bosonic double-semion state contains  quasi-electron and quasi-hole excitations with charge $\pm 1/2$. The composite fermion $f$ is a neutral fermion formed by one electron and two quasi-holes.  The composite fermion in the CFL state is known to have a dipole structure. It is hard to figure out the internal structure of the composite fermion in the CFI state from the low energy theory in Eq.~\ref{eq:cfi_low_energy}. We will only consider the low energy properties of the CFI state in this paper, which is not influenced by the internal structure of the composite fermion.

\subsection{CFI2: Exciton Condensation of  Bosons}
In the previous section we construct a CFI phase from the exciton condensation of the neutral fermions. In this section we further show that the exciton condensation of bosons in Eq.~\ref{eq:decouple_CFLs} can give a CFI phase with similar properties.  Considering the exciton condensation that $\langle b_1^\dagger b_2 \rangle \neq 0$  while $\langle f_1^\dagger f_2 \rangle = 0$ in Eq.~\ref{eq:decouple_CFLs}. From Eq.~\ref{eq:decouple_CFLs} and Eq.~\ref{eq:double-semion}, such an exciton condensation higgses $A_s-a_s$. After identifying $a_s=A_s$, the final theory of the second CFI phase is:
\begin{align}
L_{CFI_2}=L_{FS}[f_1,a_c+A_s]+L_{FS}[f_2,a_c-A_s]+L_b[a_c,A_c]
\end{align}
where,
\begin{equation}
  L_b[a_c,A_c]=-\frac{2}{4\pi}\alpha_1 d\alpha_1+\frac{2}{4\pi}\alpha_2 d\alpha_2+\frac{2}{2\pi}(\alpha_1+\alpha_2) d(A_c-a_c)
  \label{eq:double-semion_2}
\end{equation}

$L_b[a_c,A_c]$ describes a double-semion state for the two component boson $(b_1,b_2)$ which breaks the ``$S_z$'' rotation symmetry. After integrating $\alpha_1$ and $\alpha_2$, we find that there is no Chern-Simons term for $A_c$ and $a_c$. The second CFI phase can be described by:

\begin{align}
L_{CFI_2}=L_{FS}[f_1,a_c+A_s]+L_{FS}[f_2,a_c-A_s]
\label{eq:CFI2}
\end{align}
Here $f_1$ and $f_2$ form two Fermi surfaces with equal areas. $f_1$ and $f_2$ carries different charges under $A_s$. The phase is still an insulator under $A_c$ and a metal under $A_s$. The difference from the CFI1 is that the gapless charge under $A_s$ is carried by the composite fermions, instead of the monopole. In this CFI2 phase, the composite fermions can also be viewed as spinon, which is from the fractionalization of the physical exciton, similar to Ref.~\onlinecite{barkeshli2018topological,zaletel2018evidence}. But here we also have an internal $U(1)$ gauge field and as a result the physical properties are essentially different from that in the Ref.~\onlinecite{barkeshli2018topological,zaletel2018evidence}.

If we further add $\langle f_1^\dagger f_2 \rangle \neq 0$, the physical $U(1)$ symmetry corresponding to $A_s$ is broken. The resulting phase is still an insulator. It has additional goldstone mode corresponding to $\rho_s A_s ^2$. We will not go to details of this third CFI phase.

\subsection{Quantum Oscillation of the CFI Phase}
It is well-established that the CFL phase in the half-filled Landau level show quantum oscillations under additional magnetic field. In this subsection we discuss the possibility of quantum oscillations under both $B_c=dA_c$ and $B_s=d A_s$ for the CFI phases. Interestingly, we find that  if the charge $Q_s$ is fixed, the CFI phase can show quantum oscillations responding to $B_c$ satisfying the Onsager condition. If $Q_s$ is not fixed, the internal magnetic field $B_a=da$ is only partially locked to the external one: $B_a=\alpha B_c$ with $\alpha<1$. The value of $\alpha$ depends on the microscopic interaction and is fixed for each Hamiltonian.

First let us review the reason for quantum oscillations in the CFL phase.  This property of the CFL is actually quite non trivial because the Fermi surface is  neutral and only couples to the internal gauge field\footnote{In the original HLR theory, the Fermi surface also couples to the external gauge field $A$ and therefore show quantum oscillation. However, this understanding is not precise and can not be easily generalized to the CFI phase. In this paper, we insist  that the composite fermions should be neutral in the CFL phase.}. To show quantum oscillation response to $B$, the internal magnetic field $B_a=d a$ must locks to the external $B$: $B_a=B$. We do parton $C=bf$. $f$ forms a Fermi surface and couples to $a$. $b$ forms the $U(1)_2$ state and couples to $A-a$. One can easily get the free energy under $B_a$ and $B$ as:
\begin{equation}
  F=\chi_f |B_a|^2+ \chi_b |B-B_a|^2+\Delta_c |B-B_a|
  \label{eq:freeenergy_diamagnetism}
\end{equation}
Here $\chi_f$ and $\chi_b$ are diamagnetism constant discussed in Ref.~\onlinecite{sodemann2018quantum,chowdhury2018mixed}. For the $U(1)_2$ state, we has an additional term. Because the total filling is fixed, under $B-B_a$, the $U(1)_2$ state must excite quasi-electron or quasi-hole, which has an energy cost proportional to $\Delta_c>0$. Under the condition that $\Delta_c >(\chi_f-\chi_b)|B|$, the free energy is minimized by $B_a=B$. Therefore the $U(1)_2$ state can have a huge diamagnetism like ``Meissner effect'' although it is an insulator.  This is the reason why the CFL phase show quantum oscillation.

Then we discuss the response to $B_c$ for the CFI phase. Here the slave boson is in a double-semion state.  Then in the same analysis as above, we ask how it responses to $B-B_a$.  If the charge $Q_s$ under $A_s$ is fixed, then there should also be quasi electron-quasi hole excitations which cost energy linear to $|B-B_a|$. Therefore we still expect the same free energy as shown in Eq.~\ref{eq:freeenergy_diamagnetism}. As a result, $B_a=B$ is energetically favorable and the CFI phase show quantum oscillations under $B_c$ satisfying the Onsager quantization rule.  However, if $Q_s$ is not fixed, the double-semion phase can change its total density under $A_s$: $\delta \rho_s\sim B_c-B_a $ under $B_c-B_a$ and does not need to have quasi electron-quasi hole excitations.  If our microscopic model has a term $\delta \rho_s (\mathbf{r_1}) V_s(\mathbf{r_1}-\mathbf{r_2})\delta \rho_s(\mathbf{r_2})$, then there is the following energy cost $\chi_b |B-B_a|^2$ with $\chi_b \sim \int d^2 \mathbf{r} V_s(\mathbf{r})\sim V_s(\mathbf{q}=0)$. For quantum Hall bilayers with distance $d$, we have $\chi_b \sim d$. In this case there is no third term in Eq.~\ref{eq:freeenergy_diamagnetism} and we find that $B_a=\alpha B$ with $\alpha =\frac{\chi_b}{\chi_b+\chi_f}\sim \frac{d}{d+A\chi_f}$. In this situation the CFI phase can still show quantum oscillations, but it violates the Onsager rule. Under $d\rightarrow \infty$, we have $\alpha \rightarrow 1$. The partial lock of the internal magnetic field to the external one is similar to the result of Ref.~\onlinecite{chowdhury2018mixed,sodemann2018quantum}, which considers the case where the slave boson $b$ is in a weak Mott insulating phase. However, the $\alpha$ factor in the CFI phase is fixed by the microscopic Hamiltonian $V_s(\mathbf{q}=0)$ and does not have any temperature dependence.  For $\pm$ Chern bands in moir\'e systems\cite{zhang2018moir}, the valley degree of freedom plays the role of the pseudo-spin here and we expect $V_s(\mathbf q=0)=0$. As a result, there should not be any quantum oscillations if CFI phase can be found in such systems.

Last we have a short  discussion of  the responses to the layer magnetic field $B^s=d A_s$. CFI1 and CFI2 respond differently to the $B^s$. In the CFI1, because of the term $a_0 d A_s$, the composite fermion can be identified as a vortex under $A_s$. Therefore adding $B^s$ changes the fermion density: $\delta \rho^f\sim B^s$. The internal magnetic field is still zero and there is no quantum oscillation.

For CFI2, the composite fermions $f_1$ and $f_2$ carry opposite charges under $A_s$. Therefore under $B^s$, $f_1$ and $f_2$ form Jain squences with $\sigma_{xy}=\frac{n}{2 pn \pm 1}$ and $\sigma_{xy}=-\frac{n}{2 pn \pm 1}$. This is a sequence of fractional topological insulator states.

\section{CFI Phase in the Dirac Picture}
Recently a Dirac theory has been proposed to describe the CFL phase in the half-filled Landau level based on a fermionic particle-vortex duality\cite{son2015composite,wang2016half,metlitski2016particle,wang2015dual}. In the Dirac picture, the effective theory for the two decoupled CFLs in the opposite-fields quantum Hall bilayers is:

\begin{align}
  L&=\sum_{I=1,2}\left(\bar \Psi_I(i\slashed \partial+\slashed a_I)\Psi_I+\frac{1}{4\pi} A_I d a_I\right)\notag\\
  &-\mu \Psi^\dagger \tau_z \Psi   +\frac{1}{8\pi}A_1 d A_1-\frac{1}{8\pi}A_2 d A_2
\end{align}
where $I=1,2$ is the layer index. $\tau_\mu$ is the Pauli matrix on the layer degree of freedom.

In the Dirac picture, fermions $\Psi=(\Psi_1,\Psi_2)$ are vortexes and therefore its density is fixed by the external magnetic fields. For our system, one layer has an electron Fermi surface while the other layer has a hole Fermi surface. 

The phase is invariant under both time reversal and particle-hole (PH) symmetry. Time reversal acts as: $\Psi(t,\mathbf{x}) \rightarrow \tau_x \sigma_y \Psi^\dagger(-t,\mathbf{x})$, $a^0(t,\mathbf{x})\rightarrow -\tau_x a^0(-t,\mathbf{x})$, $\vec{a}(t,\mathbf{x})\rightarrow \tau_x \vec{a}_I(-t,\mathbf{x})$. PH acts as $\Psi(t,\mathbf{x}) \rightarrow i\sigma_y \Psi(-t,\mathbf{x})$, $a_I^0(t,\mathbf{x})\rightarrow a_I^0(-t,\mathbf{x})$, $\vec{a}_I(t,\mathbf{x})\rightarrow -\vec{a}_I(-t,\mathbf{x})$.

To be more convenient, we make a  redefinition: $\Psi_2 \rightarrow \Psi_2^\dagger$,  then the final effective theory is:
\begin{align}
  L&=\bar \Psi_1(i\slashed \partial+\slashed a_1)\Psi_1-\mu \Psi^\dagger_1 \Psi_1+\frac{1}{4\pi} A_1 d a_1\notag\\
  &+\bar \Psi_2(i\slashed \partial-\slashed a_2)\Psi_2-\mu \Psi^\dagger_2 \Psi_2+\frac{1}{4\pi} A_2 d a_2\notag\\
  &+\frac{1}{8\pi}A_1 d A_1-\frac{1}{8\pi}A_2 d A_2
  \label{eq:dirac_theory_full}
\end{align}

We will use Eq.~\ref{eq:dirac_theory_full} as our starting point. We still define $\Psi=(\Psi_1,\Psi_2)$ using the new definition of $\Psi_2$. The symmetries are acting in a different way. Time reversal acts as: $\Psi(t,\mathbf{x}) \rightarrow \tau_x \sigma_y \Psi(-t,\mathbf{x})$, $a^0(t,\mathbf{x})\rightarrow -\tau_x a^0(-t,\mathbf{x})$, $\vec{a}(t,\mathbf{x})\rightarrow \tau_x \vec{a}_I(-t,\mathbf{x})$. PH acts as $\Psi(t,\mathbf{x}) \rightarrow i\sigma_y \Psi(-t,\mathbf{x})$, $a_I^0(t,\mathbf{x})\rightarrow a_I^0(-t,\mathbf{x})$, $\vec{a}_I(t,\mathbf{x})\rightarrow -\vec{a}_I(-t,\mathbf{x})$. Here we have $T^2=-1$ for the time reversal. $T^2$ is actually ambiguous because we can always define a new time reversal by combining a global $U(1)_s$ transformation to it.

In this new representation, we have two electron Fermi pockets from the two layers. We still define $a_c=\frac{a_1+a_2}{2}$ and $a_s=\frac{a_1-a_2}{2}$. $A_c$ and $A_s$ follow the same convention. Then $d a_c$ carries the physical charge under $a_c$.  Inter-layer repulsion suppresses $d a_c$, similar to the Dirac theory of the conventional quantum Hall bilayers from the same magnetic field\cite{sodemann2017composite}. Then the fluctuation of $a_s$ dominates.  However, in our case $\Psi_1$ and $\Psi_2$ carry the same charge under $a_s$. As a result, $a_s$ fluctuation actually suppresses the inter-layer pairing instability. We reach the conclusion that the CFL phase described by the Dirac theory in Eq.~\ref{eq:dirac_theory_full} is stable against a small inter-layer repulsion.

Then we try to use the Dirac theory to describe the CFI1 phase in the intermediate distance proposed before in the HLR picture.  Clearly to reproduce the same insulating property as the CFI1 in the HLR picture, we need to higgs $da_c$. This can be done by the exciton condensation between the Dirac composite fermions: $\langle \Psi_1^\dagger(\mathbf k) A(\mathbf k) \Psi_2 (\mathbf k) \rangle \neq 0$, where $A(\mathbf k)$ is a $2\times 2$ matrix depending on the momentum $\mathbf{k}$.  

Here we propose the simplest order parameter $\langle \Psi^\dagger_1(\mathbf k) \Psi_2(\mathbf k)\rangle=\Delta$. One can check that it is invariant under both time reversal and the particle-hole transformation. The condensation identifies $a_2=-a_1=a$. The final theory is:

\begin{align}
  L_{CFI}&=\sum_{I=1,2}\left(\bar \Psi_I(i\slashed \partial-\slashed a)\Psi_I-\mu \Psi^\dagger_I \Psi_I \right)+(\Delta \Psi_1^\dagger \Psi_2+h.c.)\notag\\
  &-\frac{1}{2\pi} A_s d a+\frac{1}{4\pi}A_c d A_s+\frac{1}{4\pi}A_s d A_c
  \label{eq:dirac_theory_CFI}
\end{align}

The CFI in the Dirac picture is almost identical to the CFI1 phase in Eq.~\ref{eq:cfi_low_energy} from the HLR picture. The only difference is the following: each of the resulting two Fermi surfaces still has a $\pi$ Berry phase.  And from the Dirac picture it is easy to see that this phase is also particle-hole invariant. The $\pi$ Berry phase and the particle-hole symmetry can be checked numerically as is done in Ref.~\onlinecite{geraedts2016half}. 

\section{Topological Superconductor}
In this section we discuss other possibilities in the opposite-fields quantum Hall bilayers in addition to the CFI. A quite interesting phase is a topological superconductor from charge frustration induced by magnetic fields.  In the system of quantum Hall bilayers with opposite magnetic fields (or $C=\pm 1$ flat bands), the physical exciton is frustrated. In contrast, the interlayer electron-electron pair does not feel any magnetic field and can therefore has a well defined momentum. Such a bosonic Cooper pair can condense, leading to a superconductor with inter-layer pairing.

\subsection{Attractive inter-layer interaction}
For attractive interaction, we know for sure that the ground state is a superconductor.
The quantum Hall bilayers at filling $\nu_T=\frac{1}{2}+\frac{1}{2}$ with opposite fields can be mapped to the same-fields bilayers. 
 We make a particle hole transformation for the second layer $\tilde c_2=c^\dagger_2$. $\tilde c_2$ is a hole operator acting on the fully occupied $\nu=-1$ Integer Quantum Hall state of the second layer. $c_1$ and $\tilde c_2$ feel the same magnetic field as  the traditional quantum Hall bilayers. Under this mapping the inter-layer interaction changes sign.  At small $d$, it is well established that the ground state for a repulsive quantum Hall bilayers with the same fields is an exciton condensate with $\langle c_1^\dagger \tilde c_2 \rangle \neq 0$\cite{eisenstein2014exciton}.  Changing back to the language of the opposite-fields quantum Hall bilayers, this corresponds to a superconductor. At small $d$, it should be viewed as the BEC of Cooper pair. At large $d$, the superconductor can be viewed as from the inter-layer pairing of the composite fermions\cite{sodemann2017composite}.

 At small $d$,  we use a composite boson theory from flux attachment similar to the Ref.~\onlinecite{zhang2018paired}:
 \begin{equation}
   L[b_1,b_2,a]=L_b[b_1, \alpha+A_c+A_s]+L_b[b_2,\alpha+A_s-A_c]+\frac{1}{4\pi} \alpha d \alpha
 \end{equation}
where $L_b[b,\alpha]$ is the action for the boson $b$ coupled to the gauge field $\alpha$. $b_2$ is a composite boson coming from  the hole $\tilde c_2$ and therefore has the opposite charge compared to $b_1$.    For the traditional quantum Hall bilayers, condensations $\langle b_1 \rangle \neq 0$ and $\langle b_2 \rangle \neq 0$ lead to the physical exciton condensation. In our case, because of the opposite charge carried by $b_2$, the final state is physically different. Doing particle vortex dualities for  both $\langle b_1 \rangle$ and $\langle b_2 \rangle$, we get:
\begin{equation}
  L_{SC}=\frac{1}{4\pi} \alpha d \alpha +\frac{1}{2\pi}(\alpha+A_c+A_s)d a_1+\frac{1}{2\pi}(\alpha-A_c+A_s)d a_2
\end{equation}

Integrating $\alpha$ and defining $a_s=\frac{a_1+a_2}{2}$ and $a_c=\frac{a_1-a_2}{2}$, we have
\begin{equation}
  L_{SC}=-\frac{4}{4\pi}a_s d a_s+\frac{2}{2\pi} A_s d a_s+\frac{2}{2\pi}A_c d a_c-\frac{1}{4\pi}A_2 d A_2
\end{equation}
Because after  the particle-hole transformation we change the vacuum to be the fully occupied state of the second layer, we need to add a topological term  $-\frac{1}{4\pi}A_2 d A_2$.

The above action describes a  topological superconductor in the $\nu_{Kitaev}=2$ class.  At small $d$ the cooper pair is actually tightly bounded and is $s$ wave like because of the strong attraction. Therefore the superconductor is in the strong pairing limit.  A topological superconductor in the strong pairing limit is beyond the conventional BCS description.

\subsection{Possible superconductor with repulsive inter-layer interaction}
With repulsive inter-layer interaction, $s$ wave pairing is suppressed. However, higher angular momentum pairing can be possible. Superconductivity from repulsive interaction has attracted lots of studies in solid state systems\cite{kohn1965new}. Opposite-fields quantum Hall bilayers may offer a novel mechanism in this direction. In this set up, Fermi liquid is disfavored because single electron does not have a well-defined momentum. An higher angular momentum cooper pair may be a natural fate of the system even with inter-layer repulsion.

We actually have a series of superconducting states starting from the Dirac theory in Eq.~\ref{eq:dirac_theory_full}.  If we have an interlayer pairing $\langle \Psi_1(\mathbf k) A(\mathbf k) \Psi_2(-\mathbf k) \rangle \neq 0$, $a_s$ is higgsed and the fermion sector can be fully gapped. After that we have a gapless $a_c$ gauge field couples to $A_c$ as $\frac{2}{2\pi}A_c d a_c$, which describes a superconductor. For $s$ wave inter-layer pairing of the Dirac fermions, the superconductor is topologically equivalent to the $\nu_{Kitaev}=2$ superconductor in the last subsection\cite{sodemann2017composite}. For other angular momentum of the Dirac fermion pairing, we should get other  types of superconductors.  Some of them may also be good candidates with inter-layer repulsion. We leave studies of specific models to future work.

\section{Conclusion}
In summary, we propose a composite fermion insulator phase with neutral Fermi surface coupled to an internal $U(1)$ gauge field. Such a phase may be realized in a system consisting of nearly Flat Chern bands with opposite Chern numbers, which can be engineered in the  graphene or TMD moir\'e superlattices\cite{zhang2018moir,zhang2018bridging,wu2018layer}. It is interesting to search for the CFI phase and the topological superconductor phase both numerically and experimentally in this kind of system. Theoretically the CFI phase can be viewed as a simple generalization of the celebrated CFL phase in the half-filled landau level to the time reversal invariant case. As a matter of principle, a family of CFI phases with other bosonic sectors (for example, toric code) can emerge in strongly correlated spinless fermionic models.

\section{Acknowledgement}
We thank Debanjan Chowdhury, Max Metlitski and T. Senthil for useful discussions. The work is supported by NSF grant DMR-1608505 to Senthil Todadri.

\bibliographystyle{apsrev4-1}
\bibliography{CFI}

\onecolumngrid
\appendix

\end{document}